# First-principles study, fabrication and characterization of $(Zr_{0.25}Nb_{0.25}Ti_{0.25}V_{0.25})C$ high-entropy ceramic

Beilin Ye[1,a], Tongqi Wen[2,3,a], Manh Cuong Nguyen[3], Luyao Hao[2], Cai-Zhuang Wang[3,4], Yanhui Chu[1]*

[1]School of Materials Science and Engineering, South China University of Technology, Guangzhou 510641, China

[2]MOE Key Laboratory of Materials Physics and Chemistry under Extraordinary Conditions, School of Natural and Applied Sciences, Northwestern Polytechnical University, Xi'an 710072, China

[3]Ames Laboratory-USDOE, Iowa State University, Ames, IA, 50011, USA

[4]Department of Physics and Astronomy, Iowa State University, Ames, IA, 50011, USA

---

[a] Beilin Ye and Tongqi Wen contributed equally to this work.
* Corresponding author. Tel.:+86-20-82283990; fax:+86-20-82283990.
  *E-mail address:* chuyh@scut.edu.cn (Y.-H. Chu)

**Abstract**

The formation possibility of a new $(Zr_{0.25}Nb_{0.25}Ti_{0.25}V_{0.25})C$ high-entropy ceramic (ZHC-1) was first analyzed by the first-principles calculations and thermodynamical analysis and then it was successfully fabricated by hot pressing sintering technique. The first-principles calculation results showed that the mixing enthalpy of ZHC-1 was 5.526 kJ/mol and the mixing entropy of ZHC-1 was in the range of 0.693R-1.040R. The thermodynamical analysis results showed that ZHC-1 was thermodynamically stable above 959 K owing to its negative mixing Gibbs free energy. The experimental results showed that the as-prepared ZHC-1 (95.1% relative density) possessed a single rock-salt crystal structure, some interesting nanoplate-like structures and high compositional uniformity from nanoscale to microscale. By taking advantage of these unique features, compared with the initial metal carbides (ZrC, NbC, TiC and VC), it showed a relatively low thermal conductivity of $15.3 \pm 0.3$ W/(m·K) at room temperature, which was due to the presence of solid solution effects, nanoplates and porosity. Meanwhile, it exhibited the relatively high nanohardness of $30.3 \pm 0.7$ GPa and elastic modulus of $460.4 \pm 19.2$ GPa and the higher fracture toughness of $4.7 \pm 0.5$ MPa·m$^{1/2}$, which were attributed to the solid solution strengthening mechanism and nanoplate pullout and microcrack deflection toughening mechanism.

**Keywords:** High-entropy ceramics, metal carbides, first-principles calculations, mechanical performances, thermal physical properties.

**Graphical Abstract**

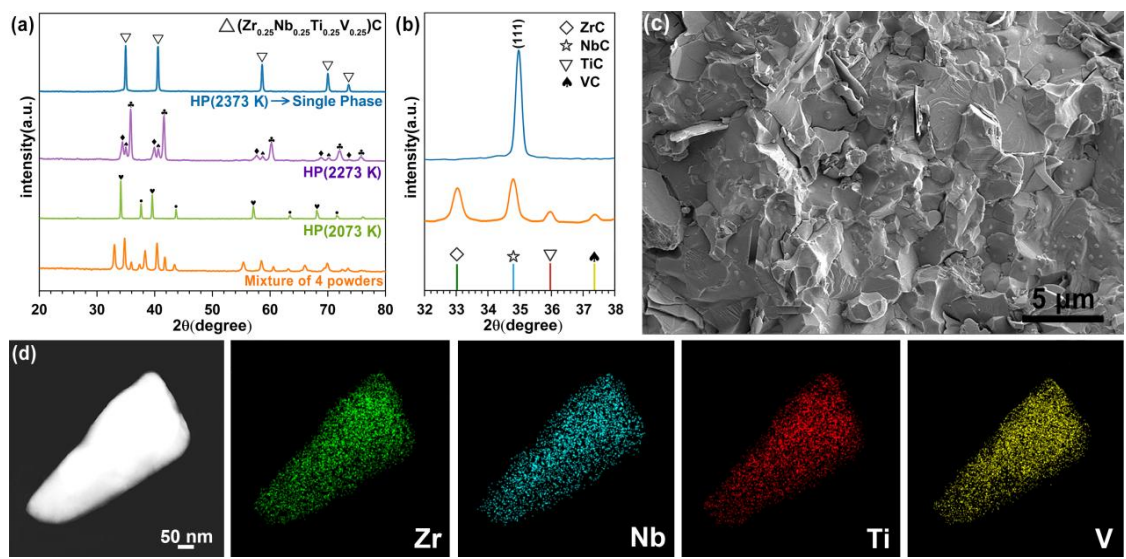

A new $(Zr_{0.25}Nb_{0.25}Ti_{0.25}V_{0.25})C$ high-entropy ceramic (ZHC-1) with a single rock-salt crystal structure of metal carbides, some interesting nanoplate-like structures and high compositional uniformity from nanoscale to microscale was fabricated and investigated in detailed and it exhibited the superior mechanical and thermal physical performances, which would endow ZHC-1 a promising candidate for extreme environmental applications.

# 1. Introduction

Crystalline high-entropy ceramics (CHCs), a new class of solid solutions needed to be composed of four or more principal metallic components in near-equiatomic ratios or at least with contents between 5 and 35 at.%, have been attracting growing attention in recent years for their unique compositions, microstructures, adjustable properties, and potential various applications [1,2]. Since 2015, the diverse CHCs, including metal oxides [3,4], diborides [5,6], nitrides [7], and carbides [8-13], have been successfully fabricated and investigated. In particular, metal carbide CHCs have recently attracted tremendous attentions for potential applications as the structural components in extreme environments owing to their excellent physical and chemical properties including high hardness and melting point and good thermal stability and corrosion resistance [14,15]. Therefore, recent research highlighted the fabrication and properties of a variety of metal carbide CHCs. At the early stage, only a couple of metal carbide CHCs, such as (Hf, Ta, Zr, Nb)C [8,9] and (Hf, Ta, Zr, Nb, Ti)C [10], were successfully synthesized and explored in experiment without the guidance from the ab-initio high-throughput computing. Recently, an entropy-forming-ability descriptor based on ab-initio high-throughput computing has been proposed to predict the formation possibility of 56 types of five-metal carbide CHCs, but only 9 types of five-metal carbide CHCs have been successfully synthesized and explored in experiment [11,12]. The as-prepared metal carbide CHCs exhibit significantly enhanced hardness and elastic modulus and reduced thermal conductivity compared with their constituent metal carbides [9-13]. Nevertheless, the comprehensive mechanical properties of the as-prepared metal carbide CHCs are still not desirable. That is to say, their hardness, elastic modulus, and fracture toughness are not able to be enhanced at the same time [13]. Therefore, extending the state of the art for more classes of metal carbide CHCs with tailored and enhanced properties is still a pressing need for the scientific community.

In this work, a new class of metal carbide CHCs, namely $(Zr_{0.25}Nb_{0.25}Ti_{0.25}V_{0.25})C$ (ZHC-1), was explored and investigated for the following aspects: (i) (Zr, Nb, Ti)C solid-solution ceramics are considered as the promising candidates for use in nuclear industrial, space and armor applications due to their low cost, high-temperature capability, high thermal and electric conductivity, low volatility, and good neutronic properties [16], and (ii) metal carbide CHCs with V element are expected to exhibit the enhanced mechanical properties since the incorporation of V element can effectively improve the mechanical properties of the metal carbide solid-solution ceramics [17,18]. In addition, the ab-initio high-throughput computing has been proposed to accurately predict the formation ability of metal carbide CHCs [11,12]. Following these pioneering works, we first analyzed the formation possibility of ZHC-1 based on the first-principles calculations and thermodynamical analysis and then successfully fabricated it by hot pressing (HP) sintering technique for the first time. The phase evolution, microstructures and mechanical and thermal physical properties of the as-prepared ZHC-1 were investigated in detail, as well as the related mechanisms. In addition, some mechanical properties were also further analyzed by first-principles calculations, such as hardness and elastic modulus. Most importantly, the resultant ZHC-1 exhibited the superior mechanical and thermal physical performances, which would endow ZHC-1 a promising candidate for extreme environmental applications.

**2. Theoretical and experimental method**

2.1 The first principles calculations

The first principles calculations based on density functional theory (DFT) were carried out using Vienna Ab-Initio Simulation Package (VASP) [19,20]. The projected-augmented-waves with the Perdew-Burke-Ernzerhof form of exchange-correlation potentials were adopted [21,22]. It was well documented that the metal carbides, such as ZrC, NbC, TiC and VC, possess two sets of face-centered cubic (FCC) sublattice [14], namely one FCC sublattice is occupied by C atoms and the other is occupied by metal atoms. As a result, it is reasonable to assume that the multi-component metal carbides would maintain the same FCC sublattice structure after forming solid solution alloys, as shown in Fig. 1, where the green balls denote C atoms while the blue ones represent metal atoms. To generate the chemical disorder in the multi-component ZHC-1, special quasi-random structure (SQS) approach was adopted [23]. In the present study, the SQS supercell with 48 atoms was constructed by using the Alloy Theoretic Automated Toolkit code [24]. For the energy calculation on the SQS structure, plane-wave basis with energy cutoff of 500 eV was used and the calculations were performed with a k-mesh grid of $2\pi \times 1/60$ Å$^{-1}$ in VASP. The electronic energy convergence criterion and the ionic force convergence criterion were $10^{-6}$ eV and 0.01 eV/Å, respectively. The unit cell shown in Fig. 1 was also used for the mechanical properties calculations on ZHC-1 and the metal carbides. In these calculations, the plane-wave cutoff energy of 800 eV was applied and the k-points Monkhorst-Pack scheme k-mesh [25] of 13 × 13 × 13 was adopted in VASP. The electronic energy convergence criterion was $10^{-8}$ eV.

The elastic constants were further calculated by the strain-stress method and then the elastic modulus calculation was performed within the Voigt-Ress-Hill approximation [26,27]. There were three independent elastic constants for cubic structure ($C_{11}$, $C_{12}$ and $C_{44}$). In general, the mechanical stability of a cubic crystal could be estimated by the generalized stability criteria which could be expressed by elastic constants: $C_{11} − C_{12} > 0$, $C_{11} > 0$, $C_{44} > 0$, $C_{11} + 2C_{12} > 0$ [28]. The elastic constants ($C_{11}$, $C_{12}$ and $C_{44}$) of ZHC-1 were 547.9 GPa, 114.9 GPa and 170.1 GPa, respectively, which satisfied the stability criteria and indicated the mechanical stability. Furthermore, the brittle-ductile behavior of materials was crucial to engineering applications and the Cauchy pressure ($C_{12} − C_{44}$) was often used to predict this behavior. The calculated Cauchy pressure ($C_{12} − C_{44}$) of ZHC-1 was negative at 0 K and 0 Pa, indicating that this system was a brittle one. On the basis of the calculated elastic constants, the "theoretical" elastic modulus and hardness were computed by the following equations [27]:

$$B = \frac{C_{11}+2C_{12}}{3} \tag{1}$$

$$G = \frac{G_V + G_R}{2} \tag{2}$$

$$G_V = \frac{3C_{44}+C_{11}-C_{12}}{5} \tag{3}$$

$$G_R = \frac{5(C_{11}-C_{12})C_{44}}{4C_{44}+3(C_{11}-C_{12})} \tag{4}$$

$$E = \frac{9BG}{3B+G} \tag{5}$$

$$H_V = 2(k^2G)^{0.585} - 3 \tag{6}$$

where $B$ is bulk modulus, $G$ is shear modulus, $G_V$ is Voigt shear modulus, $G_R$ is Reuss shear modulus, and $k$ is Pugh's modulus ratio, defined as $G/B$.

2.2 Preparation of ZHC-1

The commercially available ZrC, NbC, TiC and VC powders (99.9% purity, average particle size < 3 μm, Shanghai ChaoWei Nanotechnology Co. Ltd., Shanghai, China) were utilized as starting materials to prepare ZHC-1 by HP sintering technique in the furnace. The powders were first mixed with a ratio of 25 mol.% ZrC, 25 mol.% NbC, 25 mol.% TiC, and 25 mol.% VC, and then ball-milled for 24 h in ethanol using high-purity agate spherical media. Afterwards, they were dried, screened, and compacted into pellets of 20 mm (diameter) × 10 mm (thickness) under a uniaxial pressure of 10 MPa. Finally, these pellets were hot-pressed inside graphite dies under a biaxial pressure of 30 MPa in vacuum at 2373 K for 30 min with a heating rate of 8 K/min.

2.3 Characterization

Nanoindentation test was performed to measure the nanohardness and elastic modulus of the samples by Nano-Indenter TM XP (MTS system Corp., Minnesota, USA) system with a diamond Berkovich indenter with a tip radius of 20 nm. Before measurement, nanoindentation set-up was calibrated using a standard silica sample. Then the nanoindentation test was performed on the well-polished surface of the samples with a space of 10 μm between the two adjacent points at a constant load (8 mN). Such an operation was repeated for 50 times. The nanohardness and elastic modulus of the samples were calculated by the Oliver and Pharr method based on the estimated Poisson's ratio (0.21) from the average value of four individual metal carbides [29,30]. The microhardness of the samples was measured on their well-polished surface by the indentation technique using the microhardness tester with a Vickers indenter (HVS-30Z, Shanghai SCTMC Co. Ltd., Shanghai, China) under different loads (0.98 N, 1.96 N, 2.94 N, 4.9 N, 9.8 N, 19.6 N, 29.4 N and 49 N). Furthermore, the fracture toughness of the samples was also measured on their well-polished surface by the indentation technique using the same microhardness tester. A load of 49 N was used to generate the cracks in the sample and the fracture toughness ($K_{Ic}$) was calculated by using Antis equation [31]:

$$K_{Ic} = 0.016 \left(\frac{E}{H}\right)^{1/2} \frac{P}{c^{3/2}} \tag{7}$$

where $P$ is the applied load, $E$ is the elastic modulus, $H$ is the microhardness, and $c$ is the radial crack length (measured from center of indent). A total of 30 indentations were performed to get the average value of the fracture toughness. Thermal physical properties of the samples at room temperature were measured by the laser flash analysis (LFA-427, Netzsch, Selb, Germany) using the method introduced by Parker [32]. Three samples were tested and the final thermal physical properties were obtained by the average values.

The samples were analyzed by X-ray diffraction (XRD, X'pert PRO; PANalytical, Almelo, Netherlands), scanning electron microscopy (SEM, supra-55; Zeiss, Oberkochen, Germany) with energy dispersive spectroscopy (EDS), transmission electron microscopy (TEM, JEM 2100, JEOL, Tokyo, Japan) with EDS. The densities of the samples were measured via the Archimedes method and the relative densities were calculated via theoretical densities that were determined by the lattice constants measured by XRD.

**3. Results and discussion**

The equilibrium lattice constants of ZHC-1 and the metal carbides at 0 K from the first principles calculations are listed in Table 1. It can be found that the calculated lattice constant of ZHC-1 is 4.434 Å, which is in good agreement with the calculated value of 4.421 Å from Vegard's Law [33]. This indicates that the equilibrium lattice constant of ZHC-1 calculated by the first principles is credible. In addition, on the basis of the energies from DFT calculations for ZHC-1 and the metal carbides at 0 K after relaxation (Table 1), we can analyze the thermodynamic stability of ZHC-1 structure. The thermodynamic stability of ZHC-1 is determined by its mixing Gibbs free energy ($\Delta G_{mix}$), which can be expressed as:

$$\Delta G_{mix} = \Delta H_{mix} - T\Delta S_{mix} \qquad (8)$$

where $\Delta H_{mix}$ is the mixing enthalpy of ZHC-1, $\Delta S_{mix}$ is the mixing entropy of ZHC-1, and $T$ is the temperature. The $\Delta H_{mix}$ should be insensitive to the temperature and thus its variation can be negligible [34]. In other words, the $\Delta H_{mix}$ of ZHC-1 can be estimated by its value at 0 K. To be specific, the mixing enthalpy of ZHC-1 at 0 K and 0 Pa ($\Delta H_{mix}^{0K}$) can be calculated by the following equation:

$$\Delta H_{mix}^{0K} = E_{ZHC-1} - (E_{ZrC} + E_{NbC} + E_{TiC} + E_{VC})/4 \qquad (9)$$

where $E$ is DFT energies of the different systems after relaxation at 0 K and 0 Pa.

Using the energies from DFT calculations, the $\Delta H_{mix}^{0K}$ of ZHC-1 is calculated to be 5.526 kJ/mol, which means that the formation of the metal sublattice with four different elements is an endothermic process. Considering a system containing two different sublattices: *h* represents one sublattice with a number of sites *X*, and *k* represents another sublattice with a number of sites *Y*, the mixing entropy of the ordered structure can be defined as [35]:

$$\Delta S_{mix} = -R \left\{ \frac{X}{X+Y} \sum_{i=1}^{N_h} x_i^h \ln(x_i^h) + \frac{Y}{X+Y} \sum_{i=1}^{N_k} x_i^k \ln(x_i^k) \right\} \quad (10)$$

where *R* is the ideal gas constant, $N_h$ and $N_k$ are the elements species in the sublattice *h* and *k*, respectively, and $x_i^h$ and $x_i^k$ are the molar fractions of the constituent *i* in the sublattice *h* and *k*, respectively. In our case, ZHC-1 is composed of two different sublattices including carbon sublattice (*h*) and metal sublattice (*k*). The carbon sublattice may be occupied by carbon atoms and some vacancies, while the metal sublattice is randomly occupied by four different equal-molar metal atoms. It should be noted that the vacancies in the carbon sublattice can be considered as a second species, therefore raising the configurational entropy for the carbon sublattice to some finite value above zero. As a result, the mixing entropy of ZHC-1 results from the disordering of metal sublattices and the vacancies in the carbon sublattices. In addition, the values of *X* of carbon sublattice (*h*) and *Y* of metal sublattice (*k*) are equal to 1. As a result, the mixing entropy of ZHC-1 can be expressed as:

$$\Delta S_{mix} = -\frac{R}{2} \left\{ \sum_{i=1}^{N_h} x_i^h \ln(x_i^h) + x_v^h \ln(x_v^h) + (1 - x_v^h) \ln(1 - x_v^h) \right\} \quad (11)$$

where $x_v^h$ is the molar fraction of the vacancies in the carbon sublattice. According to Equation 11, the minimum $\Delta S_{mix}$ of ZHC-1 per mole is calculated to be 0.693R when there are no the vacancies in the carbon sublattices, while the maximum $\Delta S_{mix}$ of ZHC-1 per mole is calculated to be 1.040R when the molar fraction of the vacancies in the carbon sublattice is 0.5. By combining the calculated $\Delta H_{mix}$ of ZHC-1, the $\Delta G_{mix}$ of ZHC-1 is calculated to be negative using Equation 8 ($\Delta G_{mix} < 0$) when the temperature is above 959 K without considering the entropy contribution of the vacancies in the carbon sublattices, indicating a possible stabilized solid solution above 959 K. In addition, the parameter $\delta$, an empirical criterion for describing the comprehensive effect of the size difference in the multi-component solid solution systems, can also estimate their formation possibility, which can be expressed as follows [5,36]:

$$\delta = \sqrt{\sum_{i=1}^{n} c_i \left(1 - \frac{r_i}{\bar{r}}\right)^2} \tag{12}$$

where $n$ is the the metal carbide component species in the multi-component metal carbide solid solutions, $c_i$ is the atomic percentage of the $i$th component of the metal carbides, and here we use the lattice constants of the metal carbides as $r_i$ and $\bar{r} = \sum_{i=1}^{n} c_i r_i$. In general, a smaller $\delta$ can possibly help decrease the lattice distortions and the corresponding strain energy in systems. Recently, the parameter $\delta$ has been proposed to evaluate the formation possibility of the perovskite oxide solid solutions [2] and metal diboride solid solutions [5,36]. Although the cation-size difference for the perovskite oxide solid solutions does not correlate well with their formation possibility, a smaller $\delta$ using the lattice constants of the individual metal diboride for the metal diboride solid solutions indicates a higher formation possibility. Furthermore, it's worth noting that there is a structural similarity between the metal diboride solid solutions and the metal carbide solid solutions that the metal cations randomly occupy only one sublattice and the anions distribute on the other one, which is significantly different from the perovskite oxide solid solutions where the metal cations occupy two different sublattices and the oxygen anions occupy the other one. Generally, the values of $\delta$ for the metal diboride solid solutions that can be successfully fabricated by the sintering technique at 2273 K under a pressure of 30 MPa using the individual metal diboride as raw materials are less than 5.2% [5]. In our case, the value of $\delta$ for ZHC-1 can be calculated to be 4.59% using Equation 12, which is in the range of the values of the as-fabricated metal diboride solid solutions ($\delta \leq 5.2\%$). Therefore, it can be concluded that the formation of ZHC-1 is possible. Nevertheless, it is difficult to obtain the single-phase solid solution metal carbides due to the nature of strong covalent bonds and low self-diffusion coefficients. Synthesizing ZHC-1 needs ultrahigh temperatures in that the self-diffusion of the metal carbides typically occurs above 50% of the melting point ($T_m = 3502$ K), which is 1751 K [37]. In other words, the synthesis of the above-mentioned ZHC-1 is challenging but worth trying.

Encouraged by the theoretical analysis, we conducted a series of experiments to synthesize ZHC-1 by HP sintering technique under 30 MPa pressure at 2073 K, 2273 K and 2373 K, respectively. Fig. 2(a) displays XRD patterns of the phase evolution at different temperatures during HP sintering process. Initially, the initial four individual metal carbide phases merge to form two kinds of rock-salt structural solid solution of metal carbides at 2073 K. When the temperatures rise to 2273 K, those two solid solution phases dissolve into other three kinds of rock-salt structural solid solution of metal carbides. As the temperatures increase to 2373 K, a single rock-salt structural phase occurs, implying that ZHC-1 can be synthesized by HP sintering at 2373 K. As shown in Fig. 2(b), it is evident that the initial four individual metal carbide phases merge to form a single rock-salt crystal structure of metal carbides without other phases after the HP sintering at 2373 K. It should be noted here that the HP sintering temperature is much higher than that from the DFT prediction, which may be due to the low surface activity and self-diffusion coefficients of the starting powders and the presence of the impurities [38]. In addition, the lattice constant of ZHC-1 can be calculated to be 4.448 Å by Jade 6.5 software (Materials data incorporated, Livermore), which is in good agreement with the result from first-principles calculations (4.434 Å). Based on the calculated lattice constant from XRD analysis, the theoretical density of ZHC-1 can be calculated to be 6.654 $g/cm^3$. At the same time, the measured density of the bulk ZHC-1 sample using the Archimedes method is 6.246 $g/cm^3$. As a result, the relative density of ZHC-1 can be calculated to be 95.1%. Fig. 3 displays a typical polished surface SEM image and the corresponding EDS compositional maps of ZHC-1, from which it can be clearly seen that the compositions of ZHC-1 are perfectly uniform at the micrometer scale without evident localization of any metal elements.

TEM analysis was employed to further investigate the crystal structure and compositional homogeneity at nanoscale. Fig. 4(a) is a representative high-resolution transmission electron microscopy (HRTEM) image of ZHC-1. It clearly exhibits a periodic lattice structure, in which a set of fringes is about 0.258 nm, corresponding to the *d*-space of (111) planes of the FCC metal carbide. Moreover, the lattice parameter can be calculated to be about 4.469 Å, close to the XRD result (4.448 Å). Selected area electron diffraction (SAED) pattern along the [112] zone axis (Fig. 4(b)) indicates that ZHC-1 is the typical FCC metal carbide structure, which is in good agreement with XRD results. Fig. 4(c) presents the typical scanning transmission electron microscopy (STEM) image and EDS compositional maps at the nanometer scale. It can be observed that Zr, Nb, Ti and V elements are uniformly distributed at nanoscale and no segregation or aggregation is found throughout the scanned area. Combined with the results from XRD and SEM, it can be concluded that a single phase solid solution metal carbide ceramic, namely $(Zr_{0.25}Nb_{0.25}Ti_{0.25}V_{0.25})C$, with good compositional uniformity from nanoscale to microscale can be prepared by HP sintering at 2373 K, which verifies the rationalization of the analysis about the formation possibility of ZHC-1.

The calculated hardness and elastic modulus of ZHC-1 and four individual metal carbides in the present and previous works along with the experimental results are listed in Table 2. As for the hardness, compared with the microhardness, the nanohardness of ZHC-1 more reflects its "theoretical hardness" due to the presence of the porosity in ZHC-1. As a consequence, in order to compare with the reported nanohardness of four individual metal carbides, the nanohardness of ZHC-1 was measured and calculated using identical techniques and methods. As shown in Table 2, the present calculating results of the nanohardness and elastic modulus are in good agreement with other theoretical results. Although a remarkable discrepancy can be observed between the experimental and calculated elastic modulus except for ZHC-1, the experimental nanohardness is highly consistent with the calculated hardness. This indicates that the measured nanohardness and elastic modulus of ZHC-1 are relatively accurate. From Table 2, it can be clearly seen that the measured nanohardness and elastic modulus of ZHC-1 are 30.3 ± 0.7 GPa and 460.4 ± 19.2 GPa, respectively, which inherit the high hardness and elastic modulus of the metal carbides. It should be noted that the measured nanohardness of ZHC-1 is significantly higher than that of NbC and TiC, but is slightly lower than that of ZrC and VC. Meanwhile, the measured elastic modulus of ZHC-1 is significantly higher than that of most of individual metal carbides, such as NbC, TiC and VC, but is only comparable to that of ZrC. Nevertheless, the measured nanohardness and elastic modulus of ZHC-1 are still slightly higher than the "rule of mixture" value of nanohardness (29.0 GPa) and elastic modulus (425.8 GPa) from the average of four individual metal carbides. The increase in hardness and elastic modulus can be contributed to the solid solution strengthening effects. Similar solid solution strengthening effects can be found in (Hf, Ta)C solid solution [47] and (Zr, Nb, Ti)C solid solution [16] with the significant improvement of hardness and elastic modulus. But the solid solution strengthening effects are not significant for ZHC-1, which is due to the high mixing enthalpy of ZHC-1 (5.526 kJ/mol). In general, for most inorganic materials, there is a positive correlation between hardness and elastic modulus. At the same time, the elastic modulus is also positively related to the internal bonding. The more negative mixing enthalpy means the larger binding force between elements [37]. On the basis of the first-principles calculations, the mixing enthalpy of ZHC-1 is up to 5.526 kJ/mol at 0 K. This indicates that the interatomic bonding of ZHC-1 is relatively weak, which limits the increase of the hardness and elastic modulus. In addition, the microhardness of ZHC-1 was measured under different applied loads, and the results are displayed in Fig. 5. Obviously, the data exhibits a decrease in the microhardness of ZHC-1 as the applied load is increased. The microhardness of ZHC-1 is up to $22.5 \pm 0.6$ GPa at an applied load of 0.98 N and then it decreases to $19.1 \pm 0.5$ GPa as the applied load is increased to 49 N, which is primarily attributed to the presence of the porosity in ZHC-1. At the same time, it's worth noting that the measured microhardness of ZHC-1 is much lower than that of the measured nanohardness, which is also mainly due to the presence of the porosity in ZHC-1.

The fracture toughness of ZHC-1 was further measured by the indentation technique, and the results are listed in Table 3. Obviously, the fracture toughness of ZHC-1 is up to $4.7 \pm 0.5$ MPa·m$^{1/2}$, which is larger than that of the individual metal carbide (Table 3). Although the solid solution effects can result in the increase of the hardness and elastic modulus, it cannot lead to the increase of the toughness [13,15]. In order to reveal the toughening mechanisms, radial crack generated by Vickers indentation were analyzed. Fig. 6(a) demonstrates a representative SEM image of Vickers indentation initiated by a load of 49 N on the well-polished surface of ZHC-1. It can be seen that the radial cracks propagated from the indentation corner, in which region the highest stresses are present. Typically, the radial crack from the indentation exhibits a straight propagation path in traditional single-phase structural ceramics, while the radial cracks all exhibit a zigzag propagation path in ZHC-1. Obviously, as indicated by the red arrows in Fig. 6(b), the microcrack showed markedly deflection near some irregular plate-like grains with the lateral sizes of 2 ~ 5 μm and then propagated along the plate-like grain boundary. This process results in an increase in the propagation path and resistance of the microcrack, which can cause ZHC-1 to be toughened, suggesting a potential microcrack deflection toughening mechanism. To further investigate the toughening mechanisms, the fracture surface of ZHC-1 was also analyzed. From Fig. 6(c), it is can be clearly seen that the fracture surface of ZHC-1 exhibits a jagged morphology with the typical transgranular fracture feature. More interestingly, plenty of the irregular nanoplates are trapped between grain boundaries and protrude from the fracture surface of ZHC-1, as shown in Fig. 6(d), indicating a potential nanoplate pullout toughening mechanism. The lateral sizes of these nanoplates are in the range of 2 ~ 5 μm, in good agreement with that of the plate-like grains, and their thickness are in the range of 100 ~ 200 nm. Fig. 6(e) displays the corresponding EDS compositional maps of Fig. 6(c). It should be noted that the compositions of the nanoplates are the same with the surrounding matrix materials. This suggests that these nanoplates are *in situ* generated in ZHC-1, which can result in the good connectivity between the matrix and the nanoplates. The formation of the nanoplates may be due to the presence of a small scale of the phase separation during the solid solution formation process [3,50,51]. During the solid solution formation process, the stability of the solid solution phases will be competed by the phase separation and/or intermetallic formation [52]. Generally, the larger absolute value of the mixing enthalpy is, the more phase separation and/or intermetallic formation occurs [37]. This finally affects the microstructure of solid solutions. In our case, the mixing enthalpy of ZHC-1 is up to 5.526 kJ/mol. Therefore, the nanoplates will be prone to being *in situ* generated in ZHC-1 when the phase separation occurs during ZHC-1 formation process. Fig. 6(f) exhibits a representative SEM image of the gaps resulting from the pullout of the nanoplates. Clearly, the widths of the gaps are in the range of 100 ~ 200 nm and their lengths are in the range of 2 ~ 5 μm, in good agreement with that of the nanoplates, which confirms the presence of a potential nanoplate pullout toughening mechanism. Therefore, owing to the presence of the good connectivity between the matrix and the nanoplates, a large part of the fracture energy can be absorbed during the microcrack deflection and nanoplate pullout, which leads to the improvement in toughness to a great extent. In other words, the enhancement of the fracture toughness of ZHC-1 should be mainly contributed to the presence of the nanoplates.

Comparison of the thermal physical properties between the as-prepared ZHC-1 and the initial four individual metal carbides is shown in Table 4. The ZHC-1 shows a relatively low thermal diffusivity of $5.2 \pm 0.1$ mm$^2$/s at room temperature, and its heat capacity is measured to be about $0.49 \pm 0.04$ J/(g·K). As a result, the ZHC-1 exhibits a relatively low thermal conductivity of $15.3 \pm 0.3$ W/(m·K) at room temperature, which is much smaller than that of most of metal carbides, such as ZrC, TiC and VC, but is only comparable to that of NbC. But the measured thermal conductivity of ZHC-1 is much less than the "rule of mixture" value of thermal conductivity (27.7 W/(m·K)) from the average of four individual metal carbides. The decrease of the thermal conductivity of ZHC-1 may be due to the result of the interaction of solid solution effects, nanoplates and porosity. Firstly, the formation of ZHC-1 is a substitutional reaction among the four individual metal carbides. Those four kinds of the metal atoms are expected to randomly occupy the sites of the metal sublattice in the FCC structure. As a result, numerous substitutional atom defects can be introduced into the lattices, which will result in an increase in thermal resistance. The reason is that there is a positive correlation between substitutional atom defects and thermal resistance ($\tau^1$), which can be given by the following equation [54]:

$$\tau^{-1} = \frac{\delta^3 \omega_D^4}{4\pi v_m^3}(\varGamma_M + \varGamma_S) \tag{13}$$

where $\delta^3$ is the atomic volume, $\omega_D$ is the Debye frequency, $v_m$ is the mean velocity of the material, and $\varGamma_M$ and $\varGamma_S$ represent the scattering parameter due to mass fluctuations and strain field fluctuations, respectively. It is evident that a large number of mass and strain field fluctuations will be created around substitutional atom defects, resulting in the increase of $\varGamma_M$ and $\varGamma_S$, which will improve the thermal resistance and finally decrease the thermal conductivity of ZHC-1 [55,56]. In addition, some vacancies will also be generated in the carbon sublattice during ZHC-1 formation, which may also improve the thermal resistance and finally decrease the thermal conductivity of ZHC-1 to some extent [57]. Secondly, the presence of the nanoplates in ZHC-1 will provide a great deal of grain boundaries, which will increase the phonon scattering and thereby improve the thermal resistance. The correlation between nanoplates and thermal resistance can be expressed as [58]:

$$\tau^{-1} = \frac{v_m}{L} \qquad (14)$$

where $v_m$ is the velocity of the phonon, and $L$ is the size of the nanoplates. From Equation 14, it is evident that the presence of the nanoplates in ZHC-1 will result in the increase in thermal resistance and finally contributes to the decrease in thermal conductivity of ZHC-1. In addition to the intrinsic effects including solid solution effects and nanoplates, the presence of the porosity can also increase the phonon scattering, improve the thermal resistance and finally decrease the thermal conductivity. In our case, the relative density of ZHC-1 is only 95.1%. This indicates the presence of a certain amount of porosity in ZHC-1, which can decrease the thermal conductivity to a certain extent. In consequence, the porosity is also a factor that causes the decrease in thermal conductivity.

## 4. Conclusion

In conclusion, we had theoretically demonstrated the formation possibility of a new ZHC-1 based on the first-principles calculations and thermodynamical analysis and then successfully fabricated it by HP technique for the first time. The as-prepared ZHC-1 exhibited a single rock-salt crystal structure of metal carbides, some interesting nanoplate-like structures and high compositional uniformity from nanoscale to microscale. Therefore, compared with the initial metal carbides (ZrC, NbC, TiC and VC), it had the relatively low thermal conductivity due to the presence of solid solution effects, nanoplates and porosity. More importantly, it showed the relatively high hardness and elastic modulus and higher fracture toughness because of the presence of the solid solution strengthening mechanism and nanoplate pullout and microcrack deflection toughening mechanism. These superior performances would endow ZHC-1 a promising candidate for extreme environmental applications.

**Acknowledgements**

Y. Chu would like to acknowledge financial support from the National Key Research and Development Program of China (No. 2017YFB0703200), National Natural Science Foundation of China (No. 51802100), and Young Elite Scientists Sponsorship Program by CAST (No. 2017QNRC001). Work at Ames Laboratory was supported by the U.S. Department of Energy, Basic Energy Sciences, Division of Materials Science and Engineering. Ames Laboratory is operated for the U.S. DOE by Iowa State University under Contract No. DE-AC02-07CH11358. The computer time support came from National Energy Research Scientific Computing Center (NERSC) in Berkeley, CA.

**Author Contributions**

Y. Chu conceived and designed the experiments. Y. Chu and B. Ye performed the experiments. Y. Chu and B. Ye analyzed the data. T. Wen, M.C. Nguyen, L. Hao and C.Z. Wang performed the first-principles calculations. All authors commented on the manuscript.

**Notes**

The authors declare no competing financial interest.

Table 1. Calculated equilibrium lattice constants and DFT energies of the different systems at 0 K.

| Systems | ZHC-1 | ZrC | NbC | TiC | VC |
|---|---|---|---|---|---|
| Lattice constants (Å) | 4.434 | 4.710 | 4.482 | 4.333 | 4.157 |
| Energies (eV/atom) | -9.668 | -9.732 | -10.274 | -9.356 | -9.539 |

Table 2. Comparison of the calculated (Cal) and experimental (Exp) hardness and elastic modulus of ZHC-1 and four individual metal carbides using identical methods or techniques.

| Samples | Source | Nanohardness ($H_n$, GPa) | Elastic Modulus ($E$, GPa) | Load (mN) |
|---|---|---|---|---|
| ZrC | Exp [39,40] | 32.5 | 464.0 | 10 |
|  | Cal(current) | 23.1 | 383.0 |  |
|  | Cal [41,42] | 24.1 ~ 26.0 | 390.7 ~ 402.2 |  |
| NbC | Exp [43] | 24.5 | 406.0 | 10 |
|  | Cal(current) | 22.2 | 472.5 |  |
|  | Cal [41,44] | 24.5 ~ 25.4 | 496.3 ~ 501.9 |  |
| TiC | Exp [45] | 25.6 | 397.3 | 13 |
|  | Cal(current) | 23.5 | 425.5 |  |
|  | Cal [41,42] | 25.4 ~ 25.9 | 435.0 ~ 441.5 |  |
| VC | Exp [46] | 33.3 | 436 | 9 |
|  | Cal(current) | 28.5 | 531.8 |  |
| ZHC-1 | Exp(current) | 30.3 ± 1.2 | 460.4 ± 19.3 | 8 |
|  | Cal(current) | 26.2 | 452.9 |  |

Table 3. Comparison of the fracture toughness of ZHC-1 with three individual metal carbides.

| System | Fracture toughness ($K_{Ic}$, MPa·m$^{1/2}$) | Load (N) | Equation | Ref |
|---|---|---|---|---|
| ZrC | 2.1 ± 0.2 | 19.6 | Anstis | 44 |
| NbC | 2.9 ± 0.2 | 98 | Lawn | 48 |
| TiC | 3.3 ± 0.1 | 9.8 | Anstis | 49 |
| ZHC-1 | 4.7 ± 0.5 | 49 | Anstis | Present work |

Table 4. The measured thermal diffusivity, heat capacity and thermal conductivity of ZHC-1 and the reported four individual metal carbides (relative density: 100%) at room temperature [10,53].

| Samples | Thermal Diffusivity [mm$^2$/s] | Heat Capacity [J/(g·K)] | Thermal Conductivity [W/(m·K)] |
|---|---|---|---|
| ZrC | 15.2 | 0.37 | 33.5 |
| NbC | 6.1 | 0.57 | 16.3 |
| TiC | 8.3 | 0.19 | 22.2 |
| VC | 13.4 | 0.51 | 38.9 |
| ZHC-1 | 5.2 ± 0.1 | 0.49 ± 0.04 | 15.3 ± 0.3 |

**Figure captions**

**Fig. 1.** A simple schematic illustration of the atomic structure that does not take the lattice distortion into account of ZHC-1.

**Fig. 2.** XRD characterization of the different samples: (a) XRD patterns of the as-prepared samples at different temperatures and the mixture of four individual metal carbide powders; (b) enlargement of (a): A is ZHC-1, B is the mixture of four individual metal carbide powders, and C is the standard diffraction peaks from (111) planes of four individual metal carbides: ZrC (JCPDS card No. 65-0332), NbC (JCPDS card No. 65-7964), TiC (JCPDS card No. 65-8804), and VC (JCPDS card No. 65-8874).

**Fig. 3.** SEM image of the polished surface of ZHC-1 and the corresponding EDS compositional maps.

**Fig. 4.** TEM analysis of ZHC-1: (a) HRTEM image; (b) SAED pattern; (c) STEM image and the corresponding EDS compositional maps.

**Fig. 5.** The measured microhardness of ZHC-1 as a function of applied load.

**Fig. 6.** SEM characterization of ZHC-1: (a) SEM image of Vickers indentation initiated by a load of 49 N on the well-polished surface of ZHC-1; (b) high magnification of (a) indicates the presence of the crack deflection toughening mechanism; (c) SEM image of the fracture surface of ZHC-1; (d) high magnification of (c) indicates the presence of the nanoplate pullout toughening mechanism; (e) EDS compositional maps of (c); (f) high magnification of (c) confirms the presence of the nanoplate pullout toughening mechanism.

**Figures**

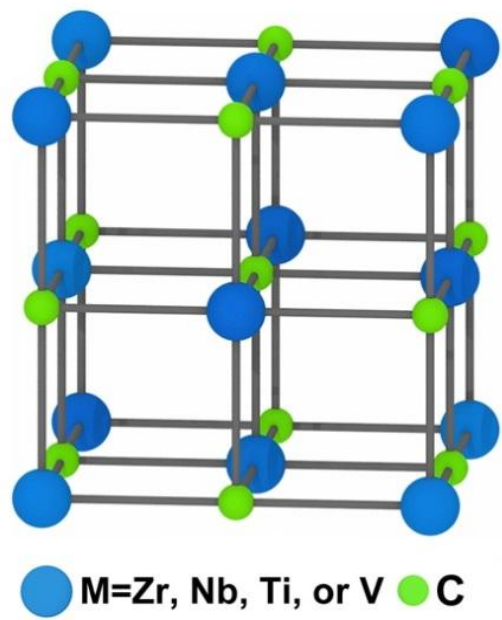

**Fig. 1.** A simple schematic illustration of the atomic structure that does not take the lattice distortion into account of ZHC-1.

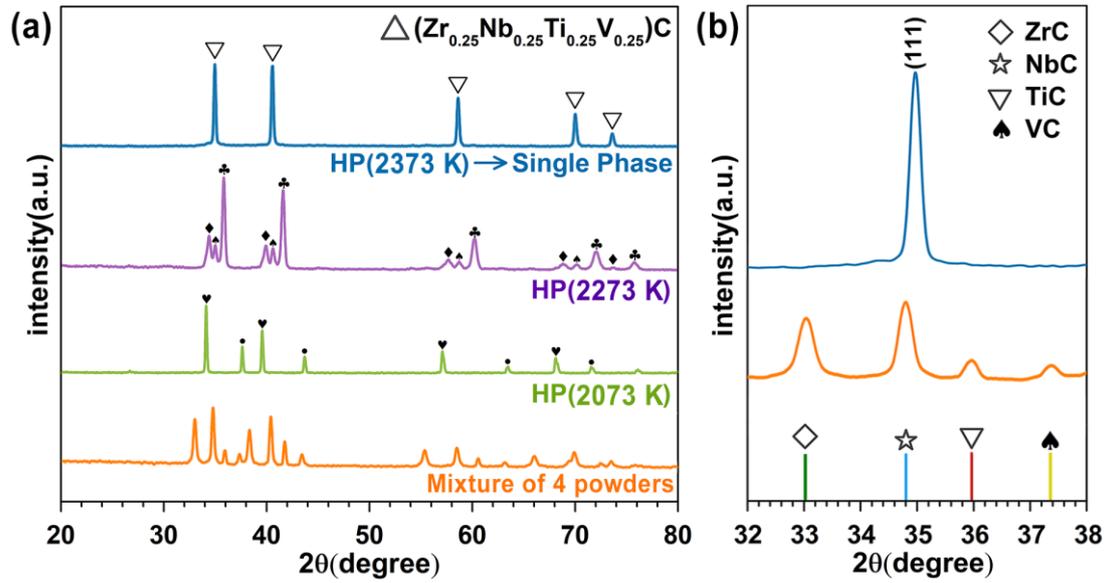

**Fig. 2.** XRD characterization of the different samples: (a) XRD patterns of the as-prepared samples at different temperatures and the mixture of four individual metal carbide powders; (b) enlargement of (a): A is ZHC-1, B is the mixture of four individual metal carbide powders, and C is the standard diffraction peaks from (111) planes of four individual metal carbides: ZrC (JCPDS card No. 65-0332), NbC (JCPDS card No. 65-7964), TiC (JCPDS card No. 65-8804), and VC (JCPDS card No. 65-8874).

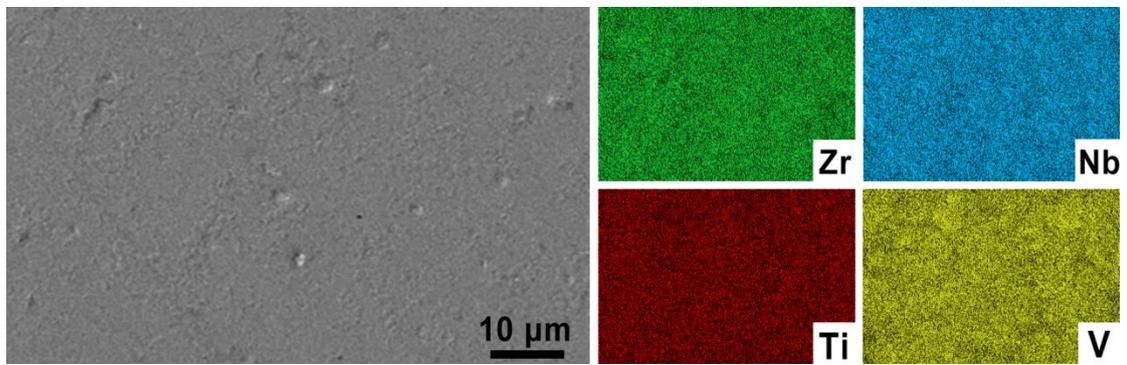

**Fig. 3.** SEM image of the polished surface of ZHC-1 and the corresponding EDS compositional maps.

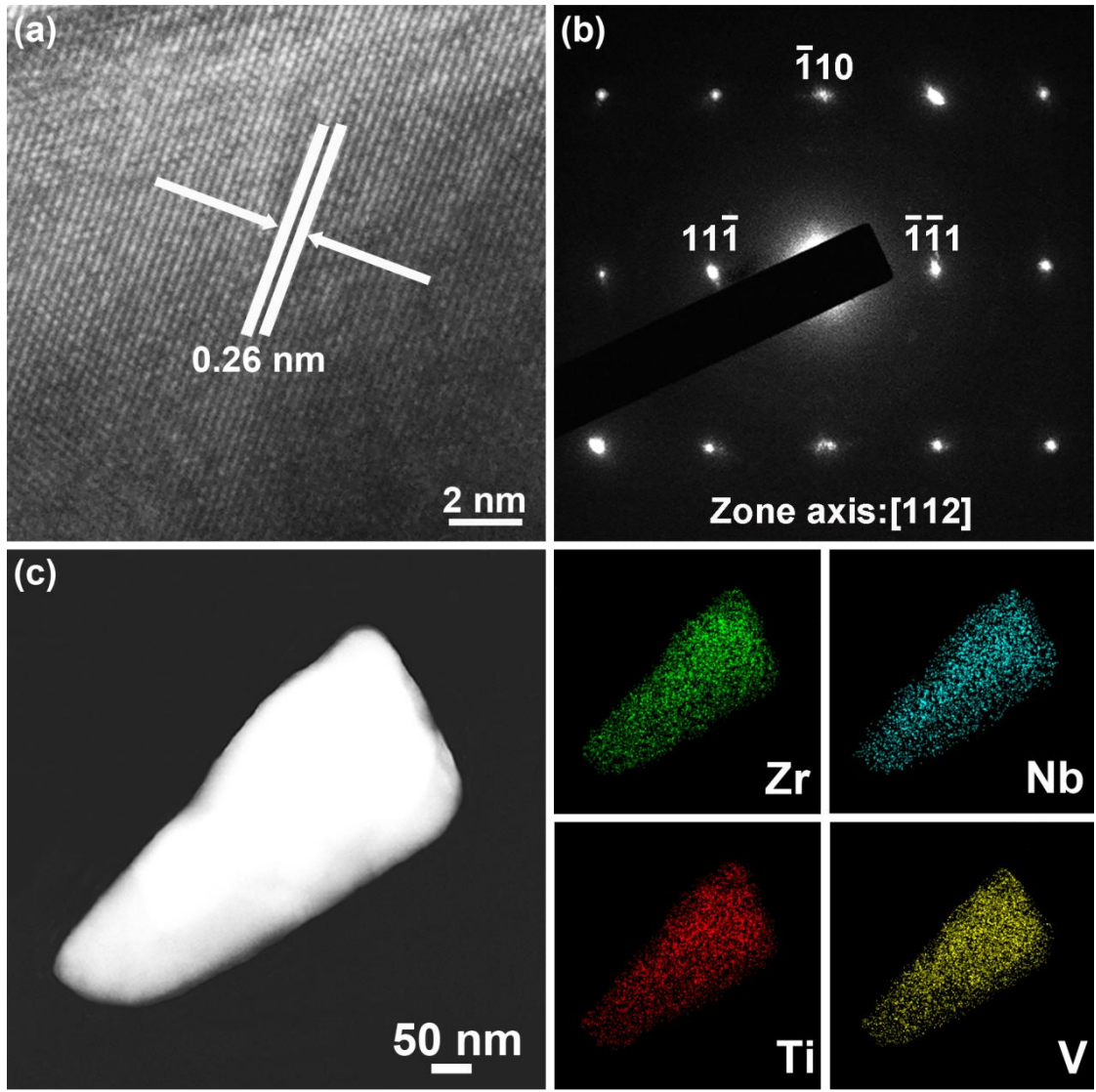

**Fig. 4.** TEM analysis of ZHC-1: (a) HRTEM image; (b) SAED pattern; (c) STEM image and the corresponding EDS compositional maps.

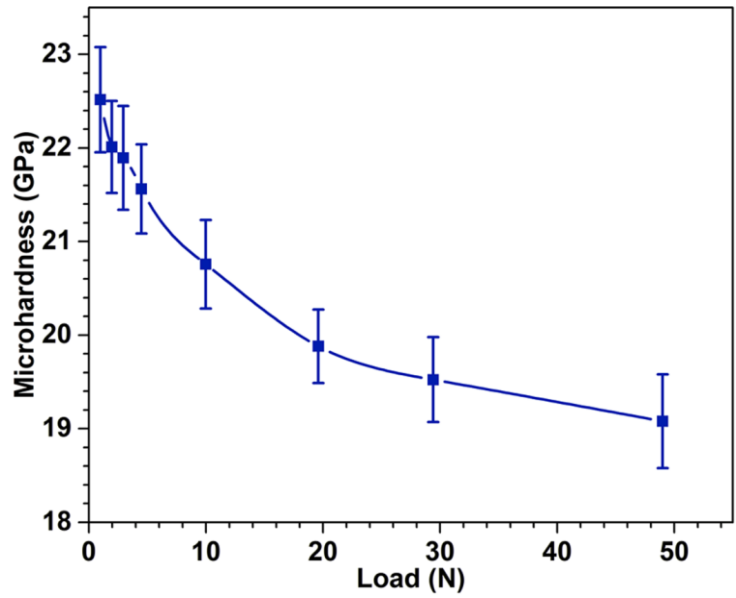

**Fig. 5.** The measured microhardness of ZHC-1 as a function of applied load.

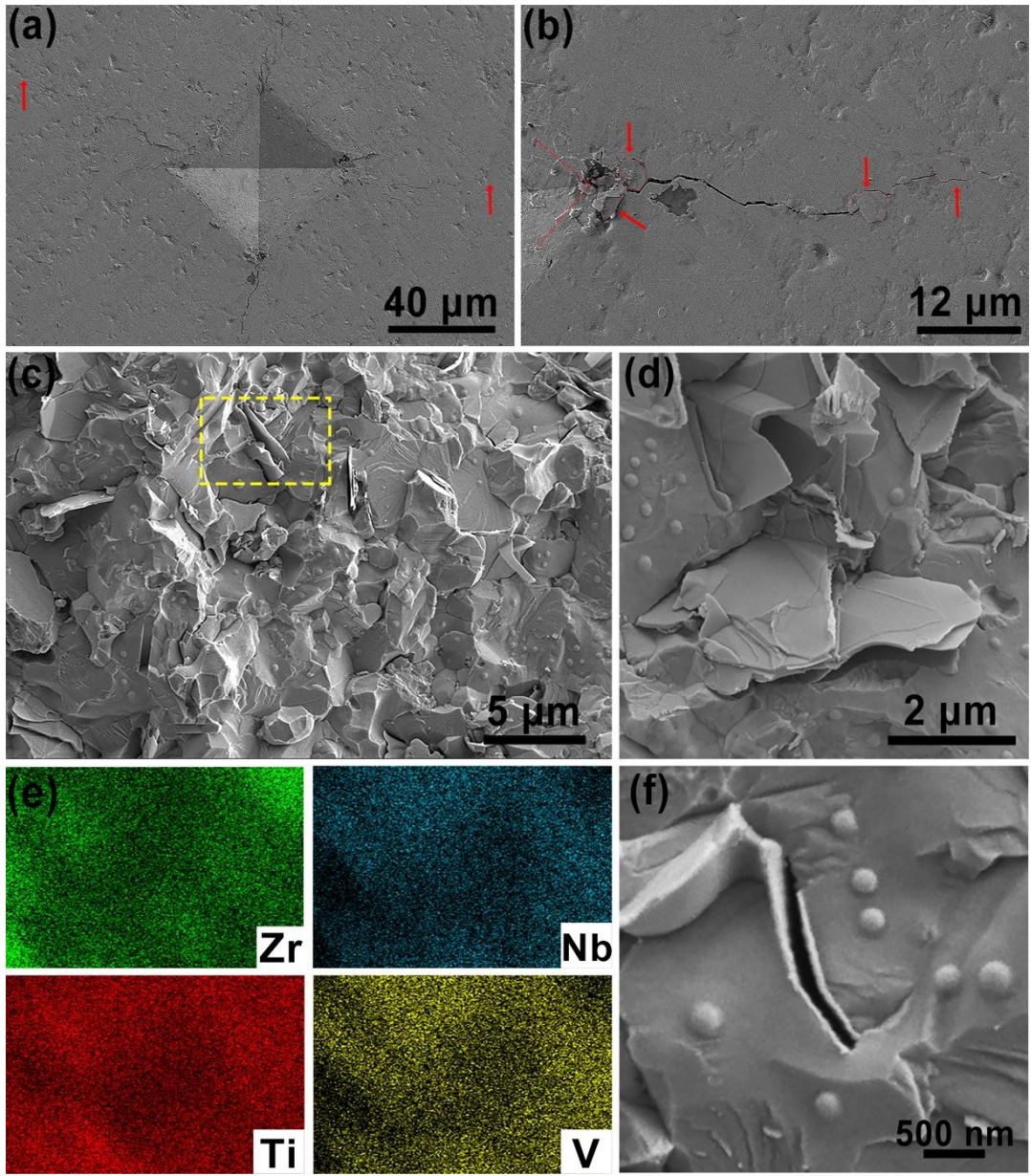

**Fig. 6.** SEM characterization of ZHC-1: (a) SEM image of Vickers indentation initiated by a load of 49 N on the well-polished surface of ZHC-1; (b) high magnification of (a) indicates the presence of the crack deflection toughening mechanism; (c) SEM image of the fracture surface of ZHC-1; (d) high magnification of (c) indicates the presence of the nanoplate pullout toughening mechanism; (e) EDS compositional maps of (c); (f) high magnification of (c) confirms the presence of the nanoplate pullout toughening mechanism.